\documentclass[conference]{IEEEtran}
\usepackage[ruled,vlined]{algorithm2e}
\usepackage{amssymb}

\ifCLASSINFOpdf
  \usepackage[pdftex]{graphicx}
\else
  \usepackage[dvips]{graphicx}
\fi
\usepackage{amsmath}
\newcommand{\hmaster}{\textit{hmaster}}
\newcommand{\hready}{\textit{hready}}
\newcommand{\hbusreq}{\textit{hbusreq}}
\newcommand{\hgrant}{\textit{hgrant}}

\newcommand{\inx}{\mathcal{X}}

\newcommand{\outy}{\mathcal{Y}}

\newcommand{\var}{\mathcal{V}}

\newcommand{\counterstrategy}{c}

\newcommand{\sys}{\mathcal{S}}

\newcommand{\env}{\mathcal{E}}

\newcommand{\true}{\textit{true}}
\newcommand{\false}{\textit{false}}

\newcommand{\always}{\textsf{\textbf{G}}}
\newcommand{\eventually}{\textsf{\textbf{F}}}
\providecommand{\next}{} 
\renewcommand{\next}{\textsf{\textbf{X}}}
\newcommand{\until}{\textsf{\textbf{U}}}

\newcommand{\boolexpr}[3]{B^{#1}_{#2}(#3)}
\newcommand{\boolxproj}[3]{B^{#1}_{\ifthenelse{\equal{#2}{}}{\inx}{#2,\inx}}(#3)}
\newcommand{\boolyproj}[3]{B^{#1}_{\ifthenelse{\equal{#2}{}}{\outy}{#2,\outy}}(#3)}
\newcommand{\boolexprvar}{\boolexpr{}{}{\var}}
\newcommand{\boolexpression}{B}

\newcommand{\req}{\textit{req}}
\newcommand{\cl}{\textit{cl}}
\newcommand{\gr}{\textit{gr}}
\newcommand{\val}{\textit{val}}

\newcommand{\candidateRefQueue}{\textit{candidateRefQueue}}

\newcommand{\refs}{\textit{refs}}
\newcommand{\new}{{new}}
\newcommand{\exploredRefs}{\textit{exploredRefs}}
\begin{document}
	
\newtheorem{example}{Example}
\newtheorem{definition}{Definition}
\newtheorem{lemma}{Lemma}
\newtheorem{proposition}{Proposition}

%
\title{Minimal Assumptions Refinement for GR(1) Specifications}

\author{\IEEEauthorblockN{Davide G. Cavezza}
\IEEEauthorblockA{Imperial College London\\
London, UK\\
Email: d.cavezza15@imperial.ac.uk}
\and
\IEEEauthorblockN{Dalal Alrajeh}
\IEEEauthorblockA{Imperial College London\\
London, UK\\
Email: dalal.alrajeh@imperial.ac.uk}
\and
\IEEEauthorblockN{Andr\'as Gy\"orgy}
\IEEEauthorblockA{DeepMind\\
London, UK\\
Email: agyorgy@google.com}}


%


\maketitle

\begin{abstract}
Reactive synthesis is concerned with finding a correct-by-construction controller from formal specifications, typically expressed in Linear Temporal Logic (LTL). The specifications describe assumptions about an environment and guarantees to be achieved by the controller operating in that environment. If a controller exists, given the assumptions, the specification is said to be realizable.

This paper focuses on finding a minimal set of assumptions that guarantee realizability in the context of counterstrategy-guided assumption refinement procedures. Specifically, we introduce the notion of minimal assumptions refinements and provide an algorithm that provably computes these with little time overhead. We show experimentally, using common benchmarks, that embedding our algorithm in state-of-the-art approaches for assumption refinement results in consistently shorter solutions than without such embedding, and allows to explore a higher number of candidate solutions. We also propose a hybrid variant for dealing with the higher sparsity of solutions in the space of minimal refinements and show that its application speeds up the identification of a solution.
\end{abstract}


%
\IEEEpeerreviewmaketitle

\section{Introduction}
Reactive synthesis is concerned with the automatic construction of  provably correct controllers from  specifications  expressed in some formal language, typically Linear Temporal Logic (LTL). Though  LTL synthesis is doubly exponential, 
more recent advances in the field  (such as \cite{Finkbeiner2016,DIppolito2010,Bloem2012}) have focused on 
synthesis problems for a subset of LTL called  \emph{generalized reactivity of rank 1} (GR(1) for short) which has a comparatively low,
polynomial time complexity \cite{Piterman2005}. 

GR(1) specifications have an assume-guarantee form: a set of assumptions about the environment and a set of guarantees for the controller. Synthesis aims is to find a controller for which all guarantees
are met in every environment that satisfies all the assumptions. If no such controller exists, the specification is said to be \textit{unrealizable}. 

Unrealizability often arises when  assumptions about the environment are too weak---allowing the 
environment  too many behaviours some of which force any controller   to violate its guarantee.
\textit{Assumptions refinement} aims at identifying  sets of \emph{sufficient} assumptions that restrict the environment  from exhibiting such violating behaviour, thus making the  specification realizable. 
In its simplest form, it defines a search problem in an intractably large search space. To deal with this, state-of-the-art approaches (e.g., \cite{Li2011b,Alur2013b,Cavezza2017,Maoz2019}) devise an incremental approach that makes use of \emph{counterstrategies}, examples of environment behaviours that force a guarantee violation. In brief,  given an unrealizable specification, a counterstrategy $\counterstrategy$ is  first computed. From this a set of alternative assumptions $\{\psi_i\}$  is generated, each of which eliminates  the counterstrategy  $\counterstrategy$ from the set of allowed environment behaviours. Each $\psi_i$ is added in turn to the original assumptions, and tested for realizability. If the new specification is still unrealizable, a new counterstrategy $\counterstrategy'$ is then computed from which a new set of alternative assumptions $\{\psi_j'\}$ is computed and so on until a set of sufficient assumptions is found.

Such incremental approaches have a number of disadvantages. First,  they are susceptible to finding solutions that are too restrictive (as they tend to over approximate violating behaviour). 
For instance, an assumption $\psi_j'$ may be sufficient alone to  eliminate both $\counterstrategy$ and $\counterstrategy'$; in this sense making $\psi_i$ in $\psi_i \wedge\psi_j'$  \emph{redundant}. 
Secondly, they  are  prone to exploring solutions that are lengthy  in the number of assumptions added. This  increases the  computation overhead  since counterstrategy computation is linear in the number of assumptions, as for Theorem 3 in \cite{Bloem2012}. It has also been argued \cite{Maoz2019} that larger sets of assumptions may negatively affect readability.
%
%
%
In this paper, we make the following contributions:
\begin{itemize}
\item We formalize the notion of redundancy of assumptions with respect to  observed counterstrategies; 
\item We provide a refinement minimization algorithm that provably removes any redundant assumption from a refinement;
\item We enhance the classical FIFO search criterion with duplicate checks to reduce the time needed to find at least one solution;
\item We show through experiments that, by integrating the minimization algorithm in counterstrategy-guided approaches, the search explores shorter solutions, in particular some that not found by existing methods alone within a given alloted time; and
\item We propose a hybrid refinement generation method that compensates for cases where 
 fewer solutions are found in the allotted time due to the minimization check.
\end{itemize}
\section{Motivating Example}
\label{sec:Motivating}

Consider the request-grant protocol described in \cite{Alur2013b}.  This protocol consists of two input variables, $\req$ and $\cl$, that the environment uses, respectively, to request access to some resource and to clear that resource; and two output variables, $\gr$ and $\val$, which, respectively, grants access to the resource and signals whether it is in a valid state.

\begin{figure}
	\centering
	\includegraphics[width=0.7\linewidth]{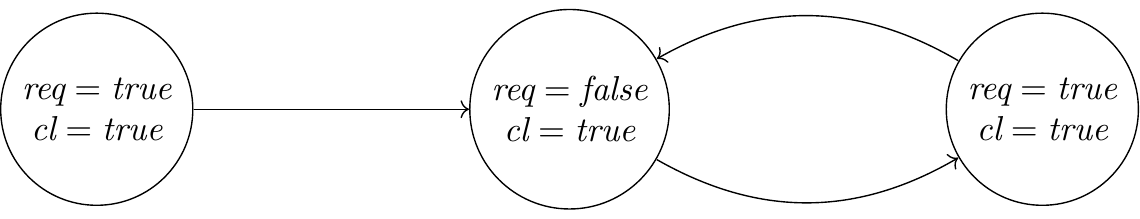}
	\caption{Simplified counterstrategy for the request-grant example. The output variables and the memory variables are omitted for clarity}
	\label{fig:counterstrategyExample2}
\end{figure}

We consider specification of request-grant protocols to be represented in GR(1), which  extends the classical Boolean logic with the operators $\always$ (``always''), $\eventually$ (``eventually''), and $\next$ (``next'');  formalized later in Section~\ref{sec:Background}. 
The  assumption $\phi^\env = \always\eventually \lnot \req$ ensures that the environment does not request access continuously. The guarantees
$\phi^\sys_1 = \always(\req \rightarrow \next\eventually \gr)$%
\footnote{Notice that this guarantee does not strictly abide by the GR(1) syntax. However, the work in \cite{Maoz2015a} shows that such formulae can be converted to pure GR(1) via the addition of auxiliary variables},
$\phi^\sys_2 = \always((\cl \lor \gr) \rightarrow \next \lnot \gr)$,
$\phi^\sys_3 = \always(\cl \rightarrow \lnot \val)$,
$\phi^\sys_4 = \always\eventually(\gr \land \val)$
guarantees that (1)  a request be eventually followed by a grant; (2) whenever a clear or a grant is active at some time step, the grant be unset at the following step; (3) whenever a clear is issued, the valid flag be unset; and (4) the environment can access valid data infinitely many times.
This specification is unrealizable since an environment that keeps the $\cl$ to $\true$ prevents any controller from fulfilling at the same time the guarantees $\phi^\sys_3$ and $\phi^\sys_4$ (Fig.~\ref{fig:counterstrategyExample2}).

Assumptions refinement procedures   search for a solution  in  a potentially infinite space of possible assumptions (or doubly exponentially large in the number of variables, if logical equivalence is taken into account). For this reason, a selection criterion, we refer to as an (inductive) \emph{bias}, must be  employed to decide which subset of assumptions to consider. Examples of such biases are given in \cite{Li2011b,Alur2013b,Cavezza2017}. 

Consider, for instance a bias based on the interpolation mechanism of \cite{Cavezza2017}. After a counterstrategy is computed (see Fig.~\ref{fig:counterstrategyExample2}), the first assumption found is $\psi = \always\eventually \lnot \cl$, which forces the environment to set $\cl$ to $\false$ infinitely many times. This is  not sufficient to achieve realizability, since an environment can force the violation of the guarantees by setting $\cl$ to $\true$ at any time following a time step where $\val$ is $\false$. This way $\phi^\sys_4$ can never be satisfied as $\gr$ and $\val$ are never satisfied together. Therefore a new counterstrategy is computed and $\psi_2 = \always (\lnot \val \rightarrow \next \lnot \cl)$ is generated. 
The refinement $\psi = \psi_1 \land \psi_2$ is sufficient, and is returned as a solution. However even the assumption $\psi_2$ alone (not found by \cite{Cavezza2017}) is sufficient, deeming $\psi_1$ redundant, and provides a weaker constraint on the environment as oppose to $\psi$.

In general, an assumptions refinement procedure has to trade off between the replacement and concatenation of assumptions when exploring potential solutions. Concatenation  typically leads to more constrained environments where realizability can be more easily achieved. Instead, replacement generates weaker refinements. This paper presents a heuristic criterion to balance between the two.
\section{Background}
\label{sec:Background}
In the following we describe the basic syntax of the GR(1) specification language, give a formal definition of counterstrategy, and define the problem of assumptions refinement.

\subsection{LTL and GR(1)}
\label{sec:LTLGR1}
LTL is an extension of propositional logic used in software engineering for describing requirements of systems \cite{Dwyer1999a,VanLamsweerde1998c,Manna1992}. 
Let $\var$ be a set of Boolean variables, and $\Sigma = 2^\var$ the set of all possible assignments/valuations to $\var$. We denote by $\Sigma^\omega$ the set of infinite sequences of elements in $\Sigma$, and for every $\pi \in \Sigma^\omega$ we denote by $\pi_i$ the $i-th$ element of the sequence. An LTL formula is an expression $\phi$ defined by the grammar (in Backus-Naur form)
$$\phi ::= p \:|\: \lnot \phi \:|\: \phi \land \phi \:|\: \next \phi \:|\: \phi \until \phi \;.$$
where $p \in \var$.
We say that a sequence $\pi \in \Sigma^\omega$ \emph{satisfies} $\phi$ (in symbols $\pi \models \phi$) if and only if one the following holds: (1) $\phi = p$ and $p \in \pi_1$, (2) $\phi = \lnot \phi'$ and $\pi \not\models \phi'$, (3) $\phi = \phi_1 \land \phi_2$ and $\pi \models \phi_1, \pi \models \phi_2$, (4) $\phi = \next \phi'$ and $\pi' = \pi_2\pi_3\dots \models \phi'$, (5) $\phi = \phi_1 \until \phi_2$, and there exists an index $k$ such that $\pi_k\pi_{k+1}\dots \models \phi_2$ and for every $i < k$ $\pi_i\pi_{i+1}\dots \models \phi_1$. The other Boolean operators ($\lor$, $\rightarrow$, $\leftrightarrow$) are defined as usual by using (1) and (2). The minimal set of temporal operators is extended with $\eventually \phi \equiv \true \until \phi$ and $\always \phi \equiv \lnot \eventually \lnot \phi$.

In this context, we are interested in a subset of LTL called \emph{GR (1)} \cite{Li2011b}. This includes formulae of the kind $\phi^\env \rightarrow \phi^\sys$, where both $\phi^\env$ and $\phi^\sys$ are conjunctions of three kinds of LTL formulae: (1) \emph{initial conditions}, purely Boolean formulae $\boolexprvar$ over the variables in $\var$ not containing any temporal operator, constraining the initial state of a system; (2) \emph{invariants}, of the form $\always \boolexpression(\var \cup \next\var)$, temporal formulae containing only an outer $\always$ operator and $\next$ operators such that no two $\next$ are nested; these formulae encode one-step transitions allowed in the system; (3) \emph{fairness conditions} of the form $\always\eventually \boolexprvar$, expressing a Boolean condition $\boolexprvar$ holding infinitely many times in a system execution. Each conjunct in $\phi^\env$ is called \emph{assumption}, and each one in $\phi^\sys$ is called \emph{guarantee}.

\subsection{GR(1) Games and Counterstrategies}
GR(1) formulae express properties of two-agent systems, where one of the agents is called \emph{environment} and the other \emph{controller}. The environment's state is characterized by a subset of variables $\inx \in \var$ called \emph{input variables}, while the controller's state is determined by the complementary subset $\outy = \var \setminus \inx$ of \emph{output variables}. The two agents compete in a game. Game states are valuations of $\var$ and transitions are consistent with assumptions and guarantees. At each step of an execution, the environment selects values for the input variables in order to fulfill the assumptions and forcing the violation of a guarantee. The controller responds by selecting a valuation of the output variables in order to satisfy the guarantees, thus taking the game to a new state. The controller synthesis problem is formalized into identifying a winning strategy for the controller in this game: a \emph{strategy} is a mapping from the history of past valuations in the game and the current input choice by the environment, to an output choice ensuring that the controller satisfy the guarantees. 
Formal definitions of games and strategies, and of the algorithm to compute controller strategies, are given in \cite{Bloem2012}.

If such a controller strategy does not exist, the GR(1) formula is said to be \emph{unrealizable}. In this case there is a strategy, called a \emph{counterstrategy}, for the environment that forces the violation of at least one guarantee \cite{Konighofer2009}. Counterstrategies  and, symmetrically to strategies, map histories of executions and the current state onto the next input choice by the environment. There are several formal definitions of counterstrategies \cite{Konighofer2009,Li2013,Cavezza2017}, all having in common its representation as a transition system whose states and/or transitions are labeled with valuations of variables. In our context, this is the only aspect of interest, therefore we give a simpler view of counterstrategies.

A \emph{counterstrategy} is a transition system $\counterstrategy = (Q_\counterstrategy, \delta_\counterstrategy, Q_0, \lambda_\counterstrategy)$, where $Q_\counterstrategy$ is a set of states, $\delta_\counterstrategy \subseteq Q_\counterstrategy \times Q_\counterstrategy$ is a set of transitions between states, $Q_0 \subseteq Q$ is a set of initial states, and $\lambda_\counterstrategy: Q \rightarrow 2^{\var \cup \mathcal{M}}$ is a labeling function for states. The labeling function assigns Boolean values to the system's variables in $\var$ and to a set of additional ``memory'' variables $\mathcal{M}$. These variables encode the history of visited states and affect the environment's next move from each state. They are added automatically by the algorithm in \cite{Konighofer2009} and are should not appear anywhere in the assumptions or the guarantees.

A \emph{run} of the counterstrategy is any infinite sequence of states $q_1q_2\dots$ with $q_1 \in Q_0$ and $(q_i,q_{i+1}) \in \delta_\counterstrategy$ for every $i \in \mathbb{N}$. Given a run, the associated \emph{play} is the infinite sequence of the labellings of its states $\pi = \lambda_\counterstrategy(q_1)\lambda_\counterstrategy(q_2)\dots$. By  definition, all of its plays satisfy the assumptions and violate at least one guarantee (in the sense of Section~\ref{sec:LTLGR1}). In the following we say that a counterstrategy $\counterstrategy$ satisfies the LTL formula $\phi$ ($\counterstrategy \models \phi$) if and only if all of its plays satisfy $\phi$. Therefore, a counterstrategy satisfies $\phi^\env$ and violates $\phi^\sys$.

\subsection{Counterstrategy-Guided Assumptions Refinement}
If a GR(1) specification $\phi^\env \rightarrow \phi^\sys$ is unrealizable, one way to fix it is to strengthen the existing assumptions so that the environment cannot force a guarantee violation under the new assumptions. A \emph{refinement} $\psi$ is a set of assumptions $\psi_1,\dots,\psi_k$ each of which is either an initial condition, an invariant, or a fairness condition. With a slight abuse of notation, we will refer to $\psi$ both as a set of assumptions $\{\psi_1,\dots,\psi_k\}$ and as their conjunction $\psi_1 \land \dots \land \psi_k$. The problem of \emph{assumptions refinement} consists in finding one or more refinements $\psi$ such that $\phi^\env \land \psi \rightarrow \phi^\sys$ is a new GR(1) formula that is realizable. We call the solutions to the assumptions refinement problem \emph{sufficient refinements}.

Usually, when no other information on the environment but the assumptions $\phi^\env$ is input to the problem, an additional goal is finding a sufficient refinement that is the \emph{weakest} possible, in the sense of being the least constraining \cite{Li2011b,Cavezza2018,Seshia2015a}. This way the controller  synthesized from the refined specification is guaranteed to meet the specification in the widest possible set of environments. Moreover, since the new assumptions are to be understood and assessed by engineers, it is desirable that they be as concise as possible.

Typical assumptions refinement approaches are based on alternating between counterstrategy computations and assumption generation, thus called \emph{counterstrategy-guided} \cite{Li2011b, Li2013, Alur2013b, Cavezza2017}. Given an unrealizable specification with assumptions $\phi$, a counterstrategy $\counterstrategy_0$ is first computed by using the algorithm in \cite{Konighofer2009}, and one or more alternative assumptions $\psi_1, \dots, \psi_h$ inconsistent with $\counterstrategy_0$ are generated, using a bias as discussed in Section~\ref{sec:Motivating}. Then, for each $\psi_i$, if $\phi \land \psi_i$ is still not sufficient, a new counterstrategy $\counterstrategy_i$ is generated, and again a set of assumptions $\psi_{i,1}, \dots \psi_{i,h'}$ inconsistent with $\counterstrategy_i$ is generated. Each of these assumptions is conjoined in turn to $\phi \land \psi_i$ and if the resulting specifications are still unrealizable, new counterstrategies are computed in the same fashion. The result is tree of refinements  that is explored in a breadth-first fashion, from shorter refinements to lengthier ones, and whose leaves are solutions to the realizability problem \cite{Cavezza2017,Alur2013b,Li2011b,Maoz2019,Li2013}. We denote this approach as \emph{breadth-first search} (BFS).

The BFS approach ensures termination, since there are only finitely many non-equivalent logical conditions and the bias ensures only a small subset of them is actually explored. At every level of the refinement tree a finite number of assumptions is generated, and along a branch of the tree the assumptions are progressively strengthened by adding new conjuncts, until either they become inconsistent or a sufficient solution is formed \cite{Cavezza2017}. This however presents a problem. Suppose the search finds a sufficient refinement $\psi = \{\psi_1, \psi_2, \dots, \psi_k\}$; suppose also that a subset of these assumptions is sufficient, that is $\psi' \subset \psi$ is also sufficient to achieve realizability. This solution would not be checked during the incremental search, since there is no backtracking along the refinement tree.

In the following, 
we will refer to collections of \emph{sets} of counterstrategies $\mathcal{C} = \{C_1,\dots,C_k\}$ . We will denote their union as $\bigcup \mathcal{C} = \bigcup_{i=1}^k C_i$.

\section{Minimal Assumptions Sets}
A \textit{sufficient refinement} is a set of assumptions such that any counterstrategy is inconsistent with their conjunction. The goal of counterstrategy-guided approaches is to increementally progress  towards a sufficient refinement by proposing at each step a candidate refinement that eliminates at least one new counterstrategy with respect to the previous candidate. If the new assumption alone removes one or more of the counterstrategies that were eliminated by the previous candidate, the corresponding assumptions generated at the previous steps may be redundant in order to form a solution. We formalize the notion of redundant assumption with respect to a set of counterstrategies.
\begin{definition}[Redundant assumptions]
	\label{def:RedundantAssumptions}
	Given a set of assumptions $\psi = \{\psi_1, \dots, \psi_n\}$ and a set of counterstrategies $C = \{\counterstrategy_1, \dots, \counterstrategy_{n}\}$, an assumption $\psi_i$ is \emph{redundant with respect to $\psi$ and $C$} if and only if for all $\counterstrategy_j \in C$, $\counterstrategy_j \not\models \bigwedge_{\psi_h \in \psi, h \ne i} \psi_h$.
\end{definition}

Notice that the definition of redundant assumption is well-formed: a counterstrategy $c \in C$ is inconsistent with a refinement $\psi = \psi_1 \land \dots \land \psi_n$ if and only if it is inconsistent with at least one $\psi_h$ alone. It is not the case that two or more assumptions are needed together to eliminate a counterstrategy $c$, so, when a redundant assumption $\psi_j$ is removed from $\psi$, there exists an assumption $\psi_h \ne \psi_j$ that eliminates $c$ alone.

In the following we will just say that an assumption $\psi_i \in \psi$ is redundant, without specifying the sets $\psi$ and $C$ when those are clear from the context.

\begin{definition}[Minimal assumptions]
	\label{def:MinimalAssumptionsSet}
	A set of assumptions $\psi$ is  \emph{minimal} with respect to a set of counterstrategies $C$ if none of its assumptions is redundant with respect to $\psi$ and $C$.
\end{definition}

In other words, an assumption $\psi_i$ in a refinement $\psi$ is redundant with respect to a set of counterstrategies $C$ if all the counterstrategies in $C$ are still eliminated when one removes $\psi_i$ from $\psi$. So, in order to construct a sufficient solution given some observed counterstrategies, $\psi_i$ may be safely deleted. A refinement without a redundant assumption is called minimal.

\section{Minimal Refinements Search}
\label{sec:MinimalRefinementsSearch}

We first describe the minimization algorithm \textsc{MinimalRefinement}. Then we present the changes to the classical FIFO refinement algorithm \cite{Alur2013b,Cavezza2017} to speed up the convergence towards a solution.

\subsection{The function \textsc{MinimalRefinement}}
\label{sec:MinimalRefinement}

The core of our proposal is the function \textsc{MinimalRefinement}, shown in Algorithm~\ref{alg:MinimalRefinement}. This function adds a new assumption $\psi'$ to an insufficient refinement $\psi$ by ensuring the resulting refinement is minimal with respect to all the counterstrategies observed so far.

Specifically, suppose an unrealizable specification $\phi^\env \rightarrow \phi^\sys$ has been refined with the refinement $\psi = \{\psi_1,\dots,\psi_n\}$, that was computed incrementally from  the counterstrategies $C = \{\counterstrategy_1,\dots,\counterstrategy_n\}$. Suppose $\psi$ is not a sufficient refinement. Hence an additional counterstrategy $\counterstrategy'$ is computed, and a new assumption $\psi'$ is generated to eliminate $\counterstrategy'$. In general, each $\psi_i \in \psi$ and $\psi'$ together may eliminate more than one counterstrategy in $C \cup \{\counterstrategy'\}$, and thus by adding $\psi'$, some $\psi_i$ may become redundant with respect to $C \cup \{\counterstrategy'\}$ and $\psi \cup \{\psi'\}$.

In order to obtain minimality in the sense of Definition~\ref{def:MinimalAssumptionsSet}, the algorithm needs to know the counterstrategies that   each assumption eliminates. The relationship between the assumptions in a refinement and the  counterstrategies they eliminate can be pictured as a bipartite graph whose vertices are the elements $\psi_i \in \psi \cup \{\psi'\}$ and $\counterstrategy_j \in C \cup \{\counterstrategy'\}$, such that there is an edge $(\psi_i, \counterstrategy_j)$ if and only if $\counterstrategy_j \not\models \psi_i$.
\begin{figure}
	\centering
	\includegraphics[width=0.6\linewidth]{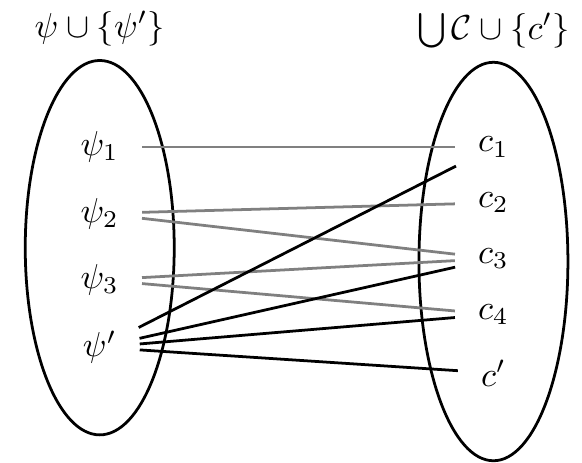}
	\caption{Bipartite graph of assumptions and counterstrategies}
	\label{fig:bipartite_graph}
\end{figure}
\begin{example}
	In Figure~\ref{fig:bipartite_graph}, $\psi_1$ eliminates $\counterstrategy_1$ only, $\psi_2$ removes $\counterstrategy_2$ and $\counterstrategy_3$, and $\psi_3$ removes $\counterstrategy_3$ and $\counterstrategy_4$. When adding $\psi'$, which removes $\counterstrategy_1$, $\counterstrategy_3$, $\counterstrategy_4$ and $\counterstrategy'$, the assumptions $\psi_1$ and $\psi_3$ become redundant. \IEEEQEDhere
\end{example}

Let the \emph{degree} of a vertex be the number of edges incident to that vertex. Then an assumption is redundant if and only if all the counterstrategies to which it is connected have a degree at least 2. Indeed, even if such an assumption were eliminated, there would be another assumption eliminating all the counterstrategies connected to it. The algorithm then scans all the $\psi_i$'s and removes all redundant assumptions based on the degree of their connected counterstrategies. The resulting refinement is minimal.
\begin{example}
	In Fig.~\ref{fig:bipartite_graph}, the addition of $\psi'$ makes $\psi_1$ and $\psi_3$ redundant. In fact, all  counterstrategies that are linked to them have degree of 2 or above. So, those two assumptions can be safely removed from the refinement. \IEEEQEDhere
\end{example}

\begin{algorithm}
	\caption{\textsc{MinimalRefinement} function}
	\label{alg:MinimalRefinement}
	\LinesNumbered
	\KwIn{$\psi = (\psi_1,\dots,\psi_k)$: insufficient refinement of cardinality $k$}
	\KwIn{$\mathcal{C}$: sequence of sets of eliminated counterstrategies with cardinality $k$}
	\KwIn{$\psi'$: assumption to be added to $\psi$}
	\KwIn{$\counterstrategy'$: a counterstrategy consistent with $\psi$ and eliminated by $\psi'$}
	\KwOut{$(\psi_\new, \mathcal{C}_\new)$: pair containing a minimal refinement $\psi \land \psi'$ and the corresponding set of eliminated counterstrategies}
	
	$C' \gets \{\counterstrategy'\}$\;	\label{alg:MinimalRefinement-line:StartComputeCPrime}	\label{alg:MinimalRefinement-line:InitCPrime}
	
	\ForEach{$\counterstrategy \in \bigcup \mathcal{C}$}{
		\If{$\counterstrategy \not\models \psi'$}{	\label{alg:MinimalRefinement-line:CSTest}
			$C' \gets C' \cup \{\counterstrategy\}$\;		\label{alg:MinimalRefinement-line:UpdateCPrime}
		}
	}
	\label{alg:MinimalRefinement-line:EndComputeCPrime}
	
	$\psi_\new \gets \psi$\;	\label{alg:MinimalRefinement-line:StartMinRefComputation} \label{alg:MinimalRefinement-line:StartInitMinRef}
	$\psi_\new$.append($\psi'$)\;
	$\mathcal{C}_\new \gets \mathcal{C}$\;
	$\mathcal{C}_\new$.append($C'$)\;		\label{alg:MinimalRefinement-line:EndInitMinRef}
	
	\ForEach{$\counterstrategy \in \bigcup \mathcal{C}_\new$}{	
		$\counterstrategy.degree$ = \textsc{CountSets}($\mathcal{C}_\new,\counterstrategy$)\;		\label{alg:MinimalRefinement-line:CountSets}
	}
	
	\For{$i = 1$ \KwTo $k$}{		\label{alg:MinimalRefinement-line:StartLoopMinRef}
		\If{$\forall \counterstrategy \in C_i \; \counterstrategy.degree \ge 2$}{	\label{alg:MinimalRefinement-line:RedundancyTest}
			$\psi_\new$.remove($\psi_i$)\;
			$\mathcal{C}_\new$.remove($C_i$)\;
			\ForEach{$\counterstrategy \in C_i$}{
				$\counterstrategy.degree \gets \counterstrategy.degree - 1$\;
			}
		}
	}		\label{alg:MinimalRefinement-line:EndLoopMinRef} \label{alg:MinimalRefinement-line:EndMinRefComputation}
	\Return{($\psi_\new,\mathcal{C}_\new$)}
	
\end{algorithm}

In order to construct the bipartite graph, the algorithm needs to determine for each pair $(\psi_i, \counterstrategy_j)$ whether $\counterstrategy_j \models \psi_i$. This is a classical model checking problem, and can be solved with standard model checking algorithms \cite{Clarke1994}. However, in this setting there is no need to perform all the model checking operations for every pair $(\psi_i, c_j)$. \textsc{MinimalRefinement} is called as part of the loop in Algorithm~\ref{alg:RefinementSearch} (see Section~\ref{sec:SearchAlgorithm}): thus at each call, part of the  graph is already constructed.

The inputs $\psi$ and $\mathcal{C}$ correspond to the bipartite graph produced by previous iterations. $\mathcal{C}$ is a collection of counterstrategy sets $C_i$ such that $\psi_i$ eliminates all the counterstrategies in $C_i$; that is, there is an edge between $\psi_i$ and each of the counterstrategies in $C_i$. $\psi'$ and $\counterstrategy'$ are the counterstrategy and the assumption produced in one iteration of the counterstrategy-guided approach (see Algorithm~\ref{alg:RefinementSearch}); hence, by hypothesis $\counterstrategy' \not\models \psi'$.

The function consists of two blocks. In the first  (lines \ref{alg:MinimalRefinement-line:StartComputeCPrime}-\ref{alg:MinimalRefinement-line:EndComputeCPrime}), the algorithm model checks all counterstrategies in $\bigcup \mathcal{C}$ against $\psi'$ and constructs the set $C' = \left\{\counterstrategy \in \bigcup \mathcal{C} \cup \{\counterstrategy'\} \:\middle|\: \counterstrategy \not\models \psi' \right\}$; by construction $\counterstrategy' \in C'$ (see line \ref{alg:MinimalRefinement-line:InitCPrime}).

The second block (lines~\ref{alg:MinimalRefinement-line:StartMinRefComputation}-\ref{alg:MinimalRefinement-line:EndMinRefComputation}) analyzes the bipartite graph and identifies the redundant assumptions. First, it builds the new refinement $\psi_\new = \psi \land \psi'$ to be minimized and the collection of counterstrategy sets $\mathcal{C}_\new$ by extending $\mathcal{C}$ accordingly with $C'$ (lines \ref{alg:MinimalRefinement-line:StartInitMinRef}-\ref{alg:MinimalRefinement-line:EndInitMinRef}). Then for every counterstrategy $\counterstrategy \in \bigcup \mathcal{C}_\new$, the function \textsc{CountSets} counts how many sets in $\mathcal{C}_\new$ contain $\counterstrategy$. This corresponds to the degree of the counterstrategy in the bipartite graph. Finally, lines~\ref{alg:MinimalRefinement-line:StartLoopMinRef}-\ref{alg:MinimalRefinement-line:EndLoopMinRef} remove any $\psi_i$ from $\psi_\new$ such that every counterstrategy in $C_i$ has a degree greater or equal to 2.

\begin{example}
	Let us consider again the motivating example in Section~\ref{sec:Motivating}. In this case, $\psi_1 = \always\eventually \lnot \cl$ eliminates the first counterstrategy generated, call it $\counterstrategy_1$. So $C_1 = \{\counterstrategy_1\}$ and $\mathcal{C} = \{C_1\}$. At the first call of \textsc{MinimalRefinement}, the bipartite graph has a single assumption vertex and a single counterstrategy vertex; so, no minimization occurs.
	
	After the second counterstrategy is produced, $\psi_2 = \always (\lnot \val \rightarrow \next \lnot \cl)$ is generated. \textsc{MinimalRefinement} checks whether $\psi_2$ is also inconsistent with $\counterstrategy_1$ and finds that it does. So $C_2 = \{\counterstrategy_1,\counterstrategy_2\}$.
	
	Now $\counterstrategy_1$ has degree 2 and $\psi_1$ is only connected to $\counterstrategy_1$. Therefore, the algorithm finds that $\psi_1$ is redundant and it removes it when constructing the new refinement $\psi_\new$.\IEEEQEDhere
\end{example}

\subsection{Search Algorithm}
\label{sec:SearchAlgorithm}
BFS approaches to assumptions refinement \cite{Alur2013b,Cavezza2017,Li2011b,Maoz2019} keep a FIFO queue of partial refinements---iteratively, the first element of the queue is extracted, and if it is not sufficient a new set of alternative further refinements is generated and appended to the queue. If a refinement of size $n$ is extracted, the appended refinements have size $n+1$, and eventually no refinement of size $n$ remains in the queue. Therefore, even if the same refinements  appears in the queue more than once, each of them is explored a finite number of times.

On the contrary using minimization could lead to the same refinement being added to the queue and explored infinitely many times. We call a refinement already explored that is added more than once to the queue a \emph{duplicate refinement}. Exploring duplicate refinements more than once can lead to useless repeated computations, hence delaying the identification of a solution, as the example below shows. 
\begin{example}
	Suppose that $\psi_1 \land \psi'_2 $ is an insufficient refinement and the bias component of the algorithm (see Section~\ref{sec:Motivating}) generates the additional assumption $\psi_3$, which makes $\psi'_2$ redundant w.r.t.~the observed counterstrategies. Suppose also that $\psi_1 \land \psi''_2$ is insufficient and again the bias generates $\psi_3$ that makes $\psi''_2$ insufficient. Then $\psi_1 \land \psi_3$ is added twice to the queue of candidate refinements. All the further refinements being extensions of $\psi_1 \land \psi_3$ will also be explored twice, hence delaying reaching any solution. \IEEEQEDhere
\end{example}
 
One simple modification to the search procedure is keeping track of the refinements (only) which have already been explored and avoiding their extension again. However, this may prevent progress towards a solution. The following example serves to argue why we should also keep track of counterstrategies eliminated by a refinement  when checking for duplicates.
\begin{example}
	Given an initial specification $ \phi^\env \rightarrow \phi^\sys$, suppose $\phi^\env \wedge \psi_1 \land \psi_2 \land \psi_3$ is a solution to the realizability problem. The following table describes an instance of  the steps within a minimal assumptions refinement search procedure.
	\begin{center}
		\begin{tabular}{ccccc}
		step \#	&  $\{(\psi_i,\bigcup \mathcal{C}_i)\}$ & $\counterstrategy$ & $\Psi'$ & \textsc{MinRef}\\
			1	& $\varnothing$ & $\counterstrategy_1$		& $\psi_1$	& $\psi_1$ \\
			&				&							& $\psi_2$	& $\psi_2 $\\
			2	& $\{(\psi_1, \counterstrategy_1)\}$		& $\counterstrategy_2$		& $\psi_2$	& $\psi_2$ \\
			3	& $\{(\psi_2, \counterstrategy_1)\}$		& $\counterstrategy_3$		& $\psi_1$	&  $\psi_1$ \\
			4	& $\{(\psi_2, \counterstrategy_1, \counterstrategy_2)\}$		& $\counterstrategy_3$		& $\psi_1$  &  $\psi_2 \land \psi_1$ \\
		\end{tabular}
	\end{center}
	The column $\{(\psi_i,\bigcup \mathcal{C}_i)\}$ represents the FIFO queue, such that the element at row $i$ is the head of the queue at step $i$. At each search step, the first element in the queue (left most) is extracted, a counterstrategy $\counterstrategy$ is computed and a set $\Psi'$ of alternative assumptions that eliminate $\counterstrategy$ is generated. The column \textsc{MinRef} shows the minimal refinement generated when adding the assumption in column $\Psi'$ to the refinement extracted from the queue given the observed counterstrategies in $\bigcup \mathcal{C}_i$ and $\counterstrategy$.
	
	At the beginning only the empty refinement is in the queue, the counterstrategy $\counterstrategy_1$ is produced and two assumptions $\psi_1$ and $\psi_2$ are generated, which are minimized and added to the queue in steps 2 and 3. Then in turn, the refinement in step 2 is extracted from the queue, $\counterstrategy_2$ is computed, $\psi_2$ is generated by the bias and after minimizing $\psi_1 \land \psi_2$ w.r.t. the counterstrategies $\counterstrategy_1$ and $\counterstrategy_2$ only $\psi_2$ remains.
	
	Notice that $\psi_2$ is generated twice: once in step 1 as a singleton refinement, and once in step 2 as the product of  minimization. Therefore, it is popped twice from positions 3 and 4. At step 4, finally $\psi_2$ is not redundant and therefore it does not disappear after minimization. If the refinement $\psi_2$ were not explored twice, the refinement $\psi_2 \land \psi_1$ would not have been added to the queue and the solution above could not be reached. The difference between the first and the second time $\psi_2$ is explored is in the counterstrategies it eliminates. \IEEEQEDhere
\end{example}

The above suggests a different way for managing the queue and checking for duplicates, as shown in Algorithm~\ref{alg:RefinementSearch}. As in existing iterative search algorithms, the function \textsc{ComputeCounterstrategy} finds a counterstrategy to an unrealizable specification, and \textsc{ApplyBias} applies a bias to compute a set of alternative assumptions that eliminate that counterstrategy. However, instead of simply storing the refinements to explore, the queue $\candidateRefQueue$ contains both the refinements and the counterstrategies from which those were generated. An additional set $\exploredRefs$ is kept in memory to contain the pairs (refinement, counterstrategies) already explored. If the pair $(\psi_i, \mathcal{C}_i)$ extracted from the queue is in $\exploredRefs$, it is no longer extended. In order to avoid the complexity of checking equivalence between sets of counterstrategies, this comparison only takes into account the number of counterstrategies in $\bigcup \mathcal{C}$ (lines 6 and 19).

\begin{algorithm}
	\caption{\textsc{RefinementSearch} function}
	\label{alg:RefinementSearch}
	\SetAlgoLined
	\LinesNumbered
	\KwIn{$\phi^\env$: set of initial assumptions}
	\KwIn{$\phi^\sys$: set of guarantees}
	\KwOut{$\refs \subseteq \{\psi \:|\: \langle \phi^\env \land \psi, \phi^\sys \rangle \text{ is realizable}\}$}
	
	$\refs \gets \varnothing$\;
	$\candidateRefQueue \gets \{\varnothing\}$\;
	$\exploredRefs \gets \varnothing$\;
	\Repeat{$\candidateRefQueue = \varnothing$}{		\label{alg:ModifiedRefinement-line:QueueLoopStart}
		$(\psi, \mathcal{C}) \gets \candidateRefQueue$.dequeue()\;		\label{alg:ModifiedRefinement-line:PopFromQueue}
		\If{$\lnot (\psi, |\bigcup \mathcal{C}|) \in \exploredRefs$}{
			\uIf{\upshape not \textsc{IsRealizable}$\langle \phi^\env \land \psi, \phi^\sys\rangle$}{		\label{alg:ModifiedRefinement-line:RealizabilityTest}
				$\counterstrategy' \gets$ \textsc{ComputeCounterstrategy}($\phi^\env \land \psi, \phi^\sys$)\;		\label{alg:ModifiedRefinement-line:ComputeCounterstrategy}
				$\Psi' \gets$ \textsc{ApplyBias}($\counterstrategy', \phi^\env \land \psi, \phi^\sys$)\;		\label{alg:ModifiedRefinement-line:GenerateAlternativeAssumptions}
				\ForEach{$\psi' \in \Psi'$}{
					$(\psi_\new, \mathcal{C}_\new) \gets$ \textsc{MinimalRefinement}($\psi, \mathcal{C}, \psi', \counterstrategy'$)\;		\label{alg:ModifiedRefinement-line:CallMinimalRefinement}
					$\candidateRefQueue$.enqueue($(\psi_\new, \mathcal{C}_\new)$)\;
					\uIf{hybrid}{
						$C' \gets $ last set in $\mathcal{C}_\new$\;
						$\candidateRefQueue$.enqueue($(\psi \cup \{\psi'\}, \mathcal{C} \cup \{C'\})$)\;
					}
				}
			}
			\uElseIf{\textsc{IsSatisfiable}($\phi^\env \land \psi$)}{		\label{alg:ModifiedRefinement-line:SatCheck}
				$\refs$.append($\psi$)\;
			}
			$\exploredRefs$.append($(\psi, |\bigcup \mathcal{C}|)$)\;
		}
	}		\label{alg:ModifiedRefinement-line:QueueLoopEnd}
	\Return{$\refs$}
\end{algorithm}
\section{Hybrid Refinement Generation}
\label{sec:Improvements}

Our minimization procedure leads to a more thorough exploration of shorter refinements within a search space compared to BFS. This allows for solutions to be found that would not have been  without minimization. This however may delays the exploration of longer refinements and hence in some cases the convergence towards a solution. This is a problem, in particular, if the specification of interest does not have short solutions. 
%
%
Moreover, an approach that solely explores minimal refinements might miss shorter solutions which may only be found by minimizing long candidate refinements, as the example below shows.

\begin{example}
	In the AMBA04 case study (see Section~\ref{sec:Experiments}) a solution is $\psi = \always (\lnot \hready \rightarrow \next \hready)$, containing only one assumption. Interpolation does not compute this assumption until some partial refinement of length 3 is given as input to \textsc{ApplyBias}. This partial refinement is found by BFS in fewer iterations than Algorithm~\ref{alg:RefinementSearch}.
\end{example}

To soften this effect, our approach can be easily hybridized with BFS. It is sufficient to add to the queue the refinements explored by BFS together with the minimal refinements generated by Algorithm~\ref{alg:MinimalRefinement}, as done in lines 14-15 of Algorithm~\ref{alg:RefinementSearch}. In this way longer candidate refinements are added to the FIFO queue in earlier iterations of the search, and longer solutions can be identified in shorter time.


\section{Experiments}
\label{sec:Experiments}

We present the results of an experiment conducted on a benchmark of six case studies of unrealizable specifications. The experiment is devised so as to show the impact of adding \textsc{MinimalRefinement} in terms of the number and the length of both the partial refinements explored and the final solutions. The experiment assumes the work shown in \cite{CavezzaA16} as a baseline. There, the authors apply a pure BFS approach to the six case studies over 24 hours and record the number of results over time, reporting also statistics of the explored refinement tree like the length. Accordingly, we executed Algorithm~\ref{alg:RefinementSearch} for 24 hours, collecting the relevant measurements on the explored refinements. The first run makes no use of the hybrid approach described in Section~\ref{sec:Improvements}. Then we repeated the experiment on all the case studies with the hybrid approach, in order to analyze its effect on the number of refinements explored and solutions found within the 24 hours. For each case study, \cite{CavezzaA16} reports results for two alternative versions of \textsc{ApplyBias}, which they call \emph{interpolation} and \emph{multivarbias}; at the time of the experiment, these are the only approaches for which data is available on the number of solutions found over a long run of the search. In our experiment, to get a stronger baseline, we selected the bias that found more solutions in BFS. 

Our implementation uses the  synthesis tool RATSY \cite{Bloem2010} to compute counterstrategies, and our Python implementation builds on the algorithm of \cite{Clarke1994} to check counterstrategies against assumptions. All approaches were executed on the same machine (Ubuntu, Intel Core i7 CPU and 16 GiB of RAM). The implementation and data are available at \cite{FMCAD19Repo}.

\subsection{Case Studies}

The six case studies are three versions of the popular AMBA-AHB protocol specification and three case studies developed in \cite{Kuvent2017} for testing Justice Violation Transition Systems (see Section~\ref{sec:Related}), called JVTS case studies for short. The AMBA-AHB specification \cite{Bloem2012,Alur2013b,Li2011b,Cimatti2008} describes the requirements of an on-chip communication protocol developed by ARM. It provides a number of masters (the environment) that can initiate a communication on a shared bus via raising an $\hbusreq_i$ signal, and an arbiter (the controller to synthesize) granting access to the bus via the signal $\hgrant_i$. An environment variable $\hready$ is set to $\true$ when the bus is ready for a switch of masters. 
The three versions of the protocol we use are the ones with respectively 2, 4 and 8 masters.

The JVTS case studies are \emph{ColorSort}, the specification of a robot sorting Lego pieces by color, \emph{GyroAspect}, a robot with self-balancing capabilities, \emph{Humanoid}, a humanoid-shaped robot. For the function \textsc{ApplyBias} we use interpolation for AMBA02, AMBA04, AMBA08 and ColorSort, and multivarbias for GyroAspect and Humanoid. The work in \cite{Kuvent2017} contains several versions of these same systems: we randomly selected one version per system and manually translated it into RATSY's syntax for adapting it to our implementation.

\subsection{Results}
\label{sec:Results}
Table~\ref{tab:ExperimentStatistics} summarizes the measurements of applying our approach (labeled as \textsc{MinimalRefinement}) by comparing it with \cite{CavezzaA16} (labeled as BFS). \emph{ExploredRefs} contains the number of non-duplicate refinements explored, that is the number of nodes for which realizability has been checked. Notice that this value is consistently higher for \textsc{MinimalRefinement}, in accordance with the fact that computing realizability and counterstrategies of smaller specifications requires sensibly less time: such gains can be observed in the average computation time despite that the length of the minimal refinements is just typically 1 to 3 assumptions less than the non-minimal ones. \emph{Min/Max/ModeLength} reports the minimum, maximum, and the mode (i.e., the value with the highest relative frequency) of the length of the explored refinements (in terms of the number of assumptions). The use of refinement minimization effectively reduces the number of assumptions in most explored refinements. This is confirmed by the row \emph{Min/Max/ModeRedundAssump}, reporting the minimum, maximum and mode of the number of redundant assumptions eliminated in each call to \textsc{MinimalRefinement}.
\emph{Sol} contains the number of distinct solutions; this count is obtained by considering just once all the refinements containing the exact same assumptions, if such refinements are found more than once during the search. However, due to time constraints no semantic check is performed on the distinct solutions: for instance, $\always\eventually(\lnot\hbusreq1)$ and $\always\eventually(\lnot\hbusreq1 \land \lnot(\hbusreq1 \land \lnot\hready))$ are considered as different assumptions, both for BFS and for \textsc{MinimalRefinement}.

As expected, our approach finds less distinct solutions than BFS (i.e., \emph{Sol} is lower); this is because minimal refinements are less constraining of the environment, and more steps are required to achieve realizability. However, the solutions found are frequently shorter: this can be read in the row \emph{Min/Max/ModeSolLength}, containing the minimum, maximum, and most frequent solution length.

To compare the solutions returned by \textsc{MinimalRefinement} and Hybrid with the ones returned by BFS, we define a notion of coverage between solutions. We say that the solution $\psi_{1}$ \emph{covers} the BFS solution $\psi_{2}$ if $\psi_{1} \subseteq \psi_{2}$; in this case, $\psi_{2}$ is bound to contain redundant assumptions not needed for realizability. For instance, in AMBA04 the solution $\always(\lnot \hready \rightarrow \next(\lnot(\lnot\hready)))$ identified by \textsc{MinimalRefinement} covers the BFS solutions $\{\always(\eventually(\lnot(\hbusreq3 \land \lnot\hmaster1))), \always(\eventually(\lnot(\lnot\hmaster1 \land \hbusreq2) \land \lnot(\hmaster0 \land \hbusreq2) \land \lnot(\hbusreq2 \land \lnot\hmaster1) \land \lnot(\hbusreq2 \land \hmaster0))), \always(\eventually(\lnot(\hmaster1 \land \hbusreq1))), \always((\lnot\hready) \rightarrow \next(\lnot(\lnot\hready)))\}$ and $\{\always(\eventually(\lnot(hbusreq3 \land \lnot\hmaster1))), \always((\lnot\hmaster1 \land \hbusreq2) \rightarrow \next(\lnot(\hbusreq2))), \always((\hmaster1 \land \hbusreq1) \rightarrow \next(\lnot(\hbusreq1))), \always((\lnot\hready) \rightarrow \next(\lnot(\lnot\hready)))\}$. 

\emph{CoveredBFSSol} shows the total number of BFS solutions covered by at least one minimal solution. \emph{SolMinimalOnly} shows the number of distinct minimal solutions that do not cover any BFS solution (and are therefore found by \textsc{MinimalRefinement} only). \emph{SolBFSOnly} shows the number of solutions that are not covered by any solution found by \textsc{MinimalRefinement}.
\emph{ExpandedRefs} shows the number of refinements that have undergone expansion, that is, the functions \textsc{ComputeCounterstrategy} and \textsc{ApplyBias} have been called on them. This number is higher in our approach, meaning that more counterstrategies could be computed within the same time.
%
%
\emph{False} contains the number of inconsistent refinements computed during the search (equivalent to the $\false$ constant), and \emph{DuplicateRefs} is the number of generated nodes discarded by our search for being duplicates  (see Section~\ref{sec:SearchAlgorithm}).


\begin{table*}[htbp]
	\caption{Summary of results of BFS, \textsc{MinimalRefinement}, and Hybrid. Check Section~\ref{sec:Results} for a description of the entries. The rows with * do not consider Hybrid}
	{\small
		\begin{center}
			\begin{tabular}{|l|l|rrrrrr|}
				\hline
				 & Statistics & \multicolumn{1}{c}{AMBA02} & \multicolumn{1}{c}{AMBA04} & \multicolumn{1}{c}{AMBA08} & \multicolumn{1}{c}{ColorSort} & \multicolumn{1}{c}{GyroAspect} &Humanoid\\ 
				\hline
				BFS & ExploredRefs & 279 & 315 & 287 & 723 & 1980 & 2734 \\
				& Min/Max/ModeLength & 0/5/5 & 0/5/5 & 0/4/4 & 0/6/6 & 0/3/3 & 0/3/3 \\
				& Sol & 84 & 48 & 10 & 28 & \textbf{896} & 150 \\ 
				& Min/Max/ModeSolLength & 1/5/4 & 3/5/4 & 4/4/4 & 3/6/5 & 2/3/3 & 1/3/3 \\
				& SolBFSOnly* & \textbf{46} & \textbf{30} & \textbf{10} & \textbf{2} & \textbf{853} & \textbf{23} \\
				& ExpandedRefs & 45 & 85 & 49 & 154 & 71 & 102 \\ 
				& False & 39 & 42 & 10 & 0 & 401 & 19 \\
				\hline
				\textsc{Minimal} & ExploredRefs & 463 & 658 & 455 & 1251 & 2139 & \textbf{3558} \\ 
				\textsc{Refinement} & Min/Max/ModeLength & \textbf{0/4/2} & \textbf{0/3/2} & 0/3/1 & \textbf{0/4/2} & \textbf{0/3/2} & \textbf{0/2/1} \\
				& Min/Max/ModeRedundAssump & 0/3/1 & 0/3/1 & 0/2/0 & 0/4/0 & 0/2/1 & 0/2/1 \\
				& Sol & 59 & 24 & 0 & 22 & 483 & 36 \\ 
				& Min/Max/ModeSolLength & \textbf{1/3/2} & \textbf{1/3/3} & N/A & \textbf{1/1/1} & \textbf{1/3/2} & \textbf{1/2/1} \\
				& CoveredBFSSol & 38 & 18 & 0 & 26 & \textbf{43} & \textbf{127} \\
				& SolMinimalOnly* & \textbf{30} & \textbf{22} & 0 & \textbf{2} & \textbf{453} & \textbf{9} \\
				& ExpandedRefs & 169 & 372 & 248 & 651 & 817 & \textbf{1236} \\ 
				& False & \textbf{11} & \textbf{2} & 0 & \textbf{0} & \textbf{173} & \textbf{5} \\
				& DuplicateRefs & 455 & 540 & 468 & 774 & 6117 & 14495\\ 
				\hline
				Hybrid & Explored refs & \textbf{1072} & \textbf{745} & \textbf{791} & \textbf{1962} & \textbf{2197} & 3494 \\
				       & Min/Max/ModeLength & 0/4/2 & 0/5/4 & 0/5/2 & 0/6/6 & 0/4/2 & 0/4/2\\
				       & Sol & \textbf{229} & \textbf{93} & \textbf{3} & \textbf{59} & 680 & \textbf{164}\\
				       & Min/Max/ModeSolLength & 1/3/2 & 1/5/4 & \textbf{1/4/4} & 1/6/6 & 1/4/2 & 1/4/2\\
				       & CoveredBFSSol & \textbf{86} & \textbf{30} & \textbf{10} & \textbf{27} & 40 & \textbf{141}\\
				       & SolHybridMinimal & 7 & 2 & 0 & 5 & 59 & 14\\
				       & SolHybridOnly & \textbf{207} & \textbf{81} & \textbf{2} & \textbf{52} & \textbf{620} & \textbf{143} \\
				       & ExpandedRefs & \textbf{248} & \textbf{398} & \textbf{234} & \textbf{667} & \textbf{1148} & 1139\\
				       & False & 11 & 10 & 3 & 0 & 272 & 20 \\
				\hline
			\end{tabular}
	\end{center}}
	\label{tab:ExperimentStatistics}
\end{table*}

In summary, although our approach produces fewer (meaningful) solutions than BFS, they are typically smaller and hence less restrictive (also note that our method explores more refinements with fewer assumptions).

Table~\ref{tab:ExperimentStatistics} also shows the effect of hybridization. \emph{SolHybridMinimal} shows the solutions  common to the hybrid approach and \textsc{MinimalRefinement}, and \emph{SolHybridOnly} the ones that do not cover any BFS solution nor  are returned by \textsc{MinimalRefinement}. In all the cases, both the number of refinements explored and the number of solutions increase. The former can be explained since more alternative assumptions (both minimal and non-minimal) are generated at every single iteration of \textsc{RefinementSearch}. The latter is due to the fact that longer solutions are generated earlier in the search. Note that the higher number of solutions allows for covering more BFS solutions, rendering the latter redundant; in AMBA08, one of the solutions covers all the 10 BFS solutions, while we were not able to find any solution by solely exploring minimal refinements. However, some of the solutions returned by Hybrid are not minimal, and therefore the length of those solutions can be higher than the ones of \textsc{MinimalRefinement}.
\section{Related Work}
\label{sec:Related}
As pointed out above, our work follows the lines of \cite{Li2011b,Alur2013b,Cavezza2017} in using counterstrategies for generating assumptions. Unlike this work, we do not propose a new bias for computing new assumptions from  individual counterstrategies. Our work, instead, focuses on a novel search strategy that mutates the refinements computed using these biases. Furthermore, our strategy for considering new refinements uses information about all the counterstrategies eliminated by that refinement instead of solely the last counterstrategy observed. 

The work in  \cite{Maoz2019} proposes an assumption refinement procedure that makes use of 
 a more efficient alternative to counterstrategies, namely Justice Violation Transition Systems (JVTSes) \cite{Kuvent2017}. A JVTS is an abstraction of counterstrategies in which cycle states and transient states are merged into ``macro-states".  Although the functions \textsc{ComputeCounterstrategy} and \textsc{ApplyBias} in Algorithm~\ref{alg:RefinementSearch} assume counterstrategies, these can an be seamlessly applied to JVTSes. This is because our \textsc{MinimalRefinement} procedure treats counterstrategies as abstract objects, only exploiting their satisfaction relation with assumptions. The main  point to consideration is the definition  of satisfaction of an LTL formula in a JVTS.

Our focus has been on synthesis from GR(1) specifications for which linear-time synthesis algorithms exist \cite{Piterman2005,Bloem2012}. A different thread of research focuses on different subsets of LTL \cite{Camacho2018,DeGiacomo2015,Patrizi2013} and reduces the synthesis  to planning. However, these approaches use explicit-state representations, and  define bounded-time versions of realizability or finite-time versions of LTL  to counter the state explosion problem. 

The definition of minimal assumptions sets is inspired by the problem of minimum set cover  \cite{Vazirani2003a}, \cite{Bazgan2005}. Given a set of elements $C$ and a collection $S$ of subsets of $C$, the minimum set cover is a subcollection of subsets $S' \subseteq S$ such that $\bigcup S' = C$ and $S'$ contains the least number of subsets needed to cover $C$. Notice that our problem of finding minimal refinements is a relaxation of this NP-hard problem, as  we are not interested in minimizing the number of subsets; rather, our notion of minimality corresponds to that of non-redundancy in \cite{Bazgan2005}. 

Redundancy and minimization in logical formulae has already been studied extensively, such as in \cite{Gottlob1993,Guthmann2017,Liberatore2005,Nguyen2014a}. The notion of redundancy we present here is substantially different from the one in the literature. Redundancy in logic is typically related to implication and entailment: a formula or a clause is redundant if another one in the knowledge base entails it. In our case, redundancy is defined with respect to the goal of eliminating a sample of counterstrategies: one assumption may not be entailed by any other in a refinement and still be redundant if all the observed counterstrategies are eliminated by the other assumptions in the refinement.
\section{Conclusions}
In this paper, we  presented a new search method for assumptions refinement to solve the unrealizability problem of GR(1) specifications. We  introduced a definition of redundancy of assumptions with respect to a set of counterstrategies and hence the concept of minimality of refinements. We proposed an algorithm that returns minimal refinements when embedded  in the classical counterstrategy-guided refinement loop. This provides a systematic way for trading off concatenation and replacement of assumptions when constructing refinements. The experiment shows that our approach explores a greater number of shorter refinements and finds shorter, non-redundant solutions. A matter for future work is exploring the effect of minimization on more recent refinement approaches such as the ones using JVTSes \cite{Maoz2019}, and the combination of minimization and weakness heuristics like the one proposed in \cite{Cavezza2018}.






\bibliographystyle{IEEEtran}
\bibliography{library}

\begin{thebibliography}{10}
\providecommand{\url}[1]{#1}
\csname url@samestyle\endcsname
\providecommand{\newblock}{\relax}
\providecommand{\bibinfo}[2]{#2}
\providecommand{\BIBentrySTDinterwordspacing}{\spaceskip=0pt\relax}
\providecommand{\BIBentryALTinterwordstretchfactor}{4}
\providecommand{\BIBentryALTinterwordspacing}{\spaceskip=\fontdimen2\font plus
\BIBentryALTinterwordstretchfactor\fontdimen3\font minus
  \fontdimen4\font\relax}
\providecommand{\BIBforeignlanguage}[2]{{%
\expandafter\ifx\csname l@#1\endcsname\relax
\typeout{** WARNING: IEEEtran.bst: No hyphenation pattern has been}%
\typeout{** loaded for the language `#1'. Using the pattern for}%
\typeout{** the default language instead.}%
\else
\language=\csname l@#1\endcsname
\fi
#2}}
\providecommand{\BIBdecl}{\relax}
\BIBdecl

\bibitem{Finkbeiner2016}
B.~Finkbeiner, ``{Synthesis of reactive systems},'' \emph{Dependable Software
  Systems Engineering}, vol.~45, pp. 72--98, 2016.

\bibitem{DIppolito2010}
N.~R. D'Ippolito, V.~Braberman, N.~Piterman, and S.~Uchitel, ``{Synthesis of
  live behaviour models},'' in \emph{Proceedings of the eighteenth ACM SIGSOFT
  international symposium on Foundations of software engineering - FSE
  '10}.\hskip 1em plus 0.5em minus 0.4em\relax New York, New York, USA: ACM
  Press, 2010, p.~77.

\bibitem{Bloem2012}
R.~Bloem, B.~Jobstmann, N.~Piterman, A.~Pnueli, and Y.~Sa'Ar, ``{Synthesis of
  Reactive(1) designs},'' \emph{Journal of Computer and System Sciences},
  vol.~78, no.~3, pp. 911--938, 2012.

\bibitem{Piterman2005}
N.~Piterman, A.~Pnueli, and Y.~Sa'ar, ``{Synthesis of Reactive(1) Designs},''
  in \emph{7th International Conference on Verification, Model Checking, and
  Abstract Interpretation (VMCAI)}, 2005, pp. 364--380.

\bibitem{Li2011b}
W.~Li, L.~Dworkin, and S.~A. Seshia, ``{Mining assumptions for synthesis},'' in
  \emph{9th ACM/IEEE International Conference on Formal Methods and Models for
  Codesign (MEMOCODE)}, 2011, pp. 43--50.

\bibitem{Alur2013b}
R.~Alur, S.~Moarref, and U.~Topcu, ``{Counter-Strategy Guided Refinement of
  GR(1) Temporal Logic Specifications},'' in \emph{Formal Methods in Computer
  Aided Design (FMCAD)}.\hskip 1em plus 0.5em minus 0.4em\relax IEEE, 2013, pp.
  26--33.

\bibitem{Cavezza2017}
D.~G. Cavezza and D.~Alrajeh, ``{Interpolation-Based GR(1) Assumptions
  Refinement},'' in \emph{Tools and Algorithms for the Construction and
  Analysis of Systems}.\hskip 1em plus 0.5em minus 0.4em\relax Springer Berlin
  Heidelberg, 2017, pp. 281--297.

\bibitem{Maoz2019}
S.~Maoz, J.~O. Ringert, and R.~Shalom, ``{Symbolic Repairs for GR(1)
  Specifications},'' \url{http://smlab.cs.tau.ac.il/syntech/repair/index.html},
  to appear in: ICSE 2019.

\bibitem{Maoz2015a}
S.~Maoz and J.~O. Ringert, ``{GR(1) synthesis for LTL specification
  patterns},'' in \emph{Proceedings of the 2015 10th Joint Meeting on
  Foundations of Software Engineering - ESEC/FSE 2015}, no.~1.\hskip 1em plus
  0.5em minus 0.4em\relax New York, New York, USA: ACM Press, 2015, pp.
  96--106.

\bibitem{Dwyer1999a}
M.~B. Dwyer, G.~S. Avrunin, and J.~C. Corbett, ``{Patterns in property
  specifications for finite-state verification},'' in \emph{Proceedings of the
  1999 International Conference on Software Engineering}, 1999, pp. 411--420.

\bibitem{VanLamsweerde1998c}
A.~{Van Lamsweerde} and L.~Willemet, ``{Inferring Declarative Requirements
  Specifications from Operational Scenarios},'' \emph{IEEE Transactions on
  Software Engineering}, vol.~24, no.~12, pp. 1089--1114, 1998.

\bibitem{Manna1992}
Z.~Manna and A.~Pnueli, \emph{The Temporal Logic of Reactive and Concurrent
  Systems}.\hskip 1em plus 0.5em minus 0.4em\relax Berlin, Heidelberg:
  Springer-Verlag, 1992.

\bibitem{Konighofer2009}
R.~Konighofer, G.~Hofferek, and R.~Bloem, ``{Debugging formal specifications
  using simple counterstrategies},'' in \emph{Formal Methods in Computer-Aided
  Design (FMCAD)}.\hskip 1em plus 0.5em minus 0.4em\relax IEEE, nov 2009, pp.
  152--159.

\bibitem{Li2013}
W.~Li, D.~Sadigh, S.~S. Sastry, and S.~A. Seshia, ``{Synthesis for
  Human-in-the-Loop Control Systems},'' in \emph{20th International Conference
  on Tools and Algorithms for the Construction and Analysis of Systems
  (TACAS)}.\hskip 1em plus 0.5em minus 0.4em\relax Springer Berlin Heidelberg,
  2014, pp. 470--484.

\bibitem{Cavezza2018}
D.~G. Cavezza, D.~Alrajeh, and A.~Gy{\"{o}}rgy, ``{A weakness measure for GR(1)
  formulae},'' in \emph{International Symposium on Formal Methods (FM)}, 2018,
  pp. 110--128.

\bibitem{Seshia2015a}
S.~A. Seshia, ``{Combining Induction, Deduction, and Structure for Verification
  and Synthesis},'' \emph{Proceedings of the IEEE}, vol. 103, no.~11, pp.
  2036--2051, nov 2015.

\bibitem{Clarke1994}
E.~Clarke, O.~Grumberg, and K.~Hamaguchi, ``{Another look at LTL model
  checking},'' in \emph{International Conference on Computer Aided Verification
  (CAV)}, vol.~71, 1994, pp. 415--427.

\bibitem{CavezzaA16}
D.~G. Cavezza and D.~Alrajeh, ``Interpolation-based {GR(1)} assumptions
  refinement,'' \emph{CoRR}, vol. abs/1611.07803, 2018.

\bibitem{Bloem2010}
R.~Bloem, A.~Cimatti, K.~Greimel, G.~Hofferek, R.~K{\"{o}}nighofer, M.~Roveri,
  V.~Schuppan, and R.~Seeber, ``{RATSY – A New Requirements Analysis Tool
  with Synthesis},'' in \emph{Computer Aided Verification}.\hskip 1em plus
  0.5em minus 0.4em\relax Springer Berlin Heidelberg, 2010, pp. 425--429.

\bibitem{FMCAD19Repo}
\BIBentryALTinterwordspacing
 [Online]. Available: \url{https://gitlab.doc.ic.ac.uk/dgc14/FMCAD19repo}
\BIBentrySTDinterwordspacing

\bibitem{Kuvent2017}
A.~Kuvent, S.~Maoz, and J.~O. Ringert, ``{A symbolic justice violations
  transition system for unrealizable GR(1) specifications},'' in
  \emph{Proceedings of the 2017 11th Joint Meeting on Foundations of Software
  Engineering - ESEC/FSE 2017}, no.~1.\hskip 1em plus 0.5em minus 0.4em\relax
  ACM Press, 2017, pp. 362--372.

\bibitem{Cimatti2008}
A.~Cimatti, M.~Roveri, V.~Schuppan, and A.~Tchaltsev, ``{Diagnostic Information
  for Realizability},'' in \emph{Verification, Model Checking, and Abstract
  Interpretation}.\hskip 1em plus 0.5em minus 0.4em\relax Berlin, Heidelberg:
  Springer Berlin Heidelberg, 2008, pp. 52--67.

\bibitem{Camacho2018}
\BIBentryALTinterwordspacing
A.~Camacho, C.~Muise, J.~A. Baier, and S.~A. McIlraith, ``{LTL Realizability
  via Safety and Reachability Games},'' in \emph{Proceedings of the
  Twenty-Seventh International Joint Conference on Artificial
  Intelligence}.\hskip 1em plus 0.5em minus 0.4em\relax California:
  International Joint Conferences on Artificial Intelligence Organization, jul
  2018, pp. 4683--4691. [Online]. Available:
  \url{https://www.ijcai.org/proceedings/2018/651}
\BIBentrySTDinterwordspacing

\bibitem{DeGiacomo2015}
G.~{De Giacomo} and M.~Y. Vardi, ``{Synthesis for LTL and LDL on finite
  traces},'' \emph{IJCAI International Joint Conference on Artificial
  Intelligence}, vol. 2015-January, no. Ijcai, pp. 1558--1564, 2015.

\bibitem{Patrizi2013}
F.~Patrizi, N.~Lipovetzky, and H.~Geffner, ``{Fair LTL synthesis for
  non-deterministic systems using strong cyclic planners},'' \emph{IJCAI
  International Joint Conference on Artificial Intelligence}, pp. 2343--2349,
  2013.

\bibitem{Kupferman2006}
O.~Kupferman, ``{Avoiding Determinization},'' in \emph{21st Annual IEEE
  Symposium on Logic in Computer Science (LICS'06)}.\hskip 1em plus 0.5em minus
  0.4em\relax IEEE, 2006, pp. 243--254.

\bibitem{Vazirani2003a}
V.~V. Vazirani, \emph{{Approximation Algorithms}}.\hskip 1em plus 0.5em minus
  0.4em\relax Springer Berlin Heidelberg, 2003.

\bibitem{Bazgan2005}
C.~Bazgan, J.~Monnot, V.~T. Paschos, and F.~Serri{\`{e}}re, ``{On the
  differential approximation of MIN SET COVER},'' \emph{Theoretical Computer
  Science}, vol. 332, no. 1-3, pp. 497--513, 2005.

\bibitem{Gottlob1993}
\BIBentryALTinterwordspacing
G.~Gottlob and C.~G. Ferm{\"{u}}ller, ``{Removing redundancy from a clause},''
  \emph{Artificial Intelligence}, vol.~61, no.~2, pp. 263--289, jun 1993.
  [Online]. Available:
  \url{https://linkinghub.elsevier.com/retrieve/pii/000437029390069N}
\BIBentrySTDinterwordspacing

\bibitem{Guthmann2017}
O.~Guthmann, O.~Strichman, and A.~Trostanetski, ``{Minimal unsatisfiable core
  extraction for SMT},'' \emph{Proceedings of the 16th Conference on Formal
  Methods in Computer-Aided Design, FMCAD 2016}, pp. 57--64, 2017.

\bibitem{Liberatore2005}
P.~Liberatore, ``{Redundancy in logic I: CNF propositional formulae},''
  \emph{Artificial Intelligence}, vol. 163, no.~2, pp. 203--232, 2005.

\bibitem{Nguyen2014a}
\BIBentryALTinterwordspacing
T.~Nguyen, D.~Kapur, W.~Weimer, and S.~Forrest, ``{Using Dynamic Analysis to
  Generate Disjunctive Invariants},'' pp. 608--619, 2014. [Online]. Available:
  \url{http://doi.acm.org/10.1145/2568225.2568275}
\BIBentrySTDinterwordspacing

\bibitem{Clarke2000}
E.~M. Clarke, Jr., O.~Grumberg, and D.~A. Peled, \emph{Model Checking}.\hskip
  1em plus 0.5em minus 0.4em\relax MIT Press, 1999.

\bibitem{Browne1997}
A.~Browne, E.~M. Clarke, S.~Jha, D.~E. Long, and W.~Marrero, ``{An improved
  algorithm for the evaluation of fixpoint expressions},'' \emph{Theoretical
  Computer Science}, vol. 178, no. 1-2, pp. 237--255, 1997.

\end{thebibliography}
%
%
%
\newpage
\appendix
\subsection{Algorithm Correctness}

We can prove the following three propositions, which formalize the above discussion.

\begin{proposition}[Coverage of all counterstrategies]
\label{prop:CounterstrategyCover}
	Suppose: (1) for every $\psi_i \in \psi$, $C_i \in \mathcal{C}$ and $\counterstrategy \in \bigcup \mathcal{C}$, $\counterstrategy \in C_i$ if and only if $\counterstrategy \not\models \psi_i$; (2) $\counterstrategy' \not\models \psi'$; (3) $\counterstrategy' \models \psi$.
	
	Then $\bigcup \mathcal{C}_\new = \bigcup \mathcal{C} \cup \{\counterstrategy'\}$.
\end{proposition}

\begin{proposition}[\textsc{MinimalRefinement} invariant]
	Under the same hypotheses on the input as in Proposition~\ref{prop:CounterstrategyCover}, $(\psi_\new,\mathcal{C}_\new)$ satisfies hypothesis (1) by replacing $\psi$ with $\psi_\new$ and $C$ with $\mathcal{C}_\new$.
\end{proposition}
\begin{IEEEproof}
	 This can be easily concluded by observing that lines~\ref{alg:MinimalRefinement-line:StartComputeCPrime}-\ref{alg:MinimalRefinement-line:EndComputeCPrime} populate $C'$ with each and every $\counterstrategy \in \bigcup \mathcal{C}_\new$ such that $\counterstrategy \not\models \psi'$; the subsequent lines that add or remove elements from $\psi_\new$ and $\mathcal{C}_\new$ are always paired. Therefore, the $j$-th element of $\mathcal{C}_\new$ contains each and every counterstrategy in $\bigcup \mathcal{C}_\new$ that is inconsistent with the $j$-th element of $\psi_\new$.
\end{IEEEproof}

\begin{proposition}[Minimality of output]
\label{prop:Minimality}
	Under the same hypotheses on the input as in Proposition~\ref{prop:CounterstrategyCover}, $\psi_\new$ is minimal with respect to $\bigcup \mathcal{C}_\new$.
\end{proposition}

This result is based on the following lemmas:
\begin{lemma}
\label{lem:BipartiteGraph}
	Let a refinement $\psi$ and a collection $C$ of counterstrategy sets fulfil hypothesis (1) of Proposition~\ref{prop:Minimality}.
	
	Let the bipartite graph $(V, E)$ be such that $V = \psi \cup \bigcup \mathcal{C}$ and $E = \left\{(\psi_i, \counterstrategy_j) \in \psi \times \bigcup \mathcal{C} \:\middle|\: \counterstrategy_j \not\models \psi_i\right\}$.
	
	Then $\psi_i$ is redundant with respect to $\psi$ and $\bigcup \mathcal{C}$ if and only if for every $\counterstrategy_j$ such that $(\psi_i, \counterstrategy_j) \in E$, the degree of $\counterstrategy_j$ is greater or equal to 2.
\end{lemma}
\begin{IEEEproof}
	Given how the graph is defined, the degree of a counterstrategy is the number of assumptions that are inconsistent with it. So, $\deg(\counterstrategy_j) \ge 2$ if and only if more than two assumptions or more eliminate it. Hence, there exists $\psi_h \ne \psi_i$ such that $\psi_h \not\models \counterstrategy_j$ for any $\counterstrategy_j$ connected to $\psi_i$. By Definition~\ref{def:RedundantAssumptions} this means $\psi_i$ is redundant.
\end{IEEEproof}

\begin{lemma}[Monotonicity w.r.t. $\psi$ of non-redundancy]
\label{lem:MonotonicityNonRedundancy}
	If $\psi_i$ is non-redundant with respect to $\psi$ and $C$, it is non-redundant w.r.t. $\mu$ and $C$ for any $\mu \subseteq \psi$.
\end{lemma}
\begin{IEEEproof}
	The intuition is trivial. If $\psi_i$ is the only assumption removing some counterstrategy $\counterstrategy_i$, then by removing some of the other assumptions in $\psi$ it will still be the only assumption removing $\counterstrategy_i$.

	If $\psi_i$ is non-redundant, by negating Definition~\ref{def:RedundantAssumptions} there exists a counterstrategy $\counterstrategy_i \in C_i$ such that $\counterstrategy_i \models \bigvee_{\psi_h \in \psi, h \ne i} \psi_h$. Since $\mu$ is a subset of $\psi$, it also holds that $\counterstrategy_i \models \bigvee_{\psi_h \in \mu, h \ne i} \psi_h$. Therefore, $\psi_i$ is non-redundant with respect to $\mu$ and $C$.
\end{IEEEproof}

Now let us prove Proposition~\ref{prop:Minimality}.
\begin{IEEEproof}\emph{(of Proposition~\ref{prop:Minimality})}
	To prove minimality, we use a loop invariant over the loop on lines~\ref{alg:MinimalRefinement-line:StartLoopMinRef}-\ref{alg:MinimalRefinement-line:EndLoopMinRef}. Let $\psi_\new^{[i]}$ be the state of $\psi_\new$ at the end of the $i$-th iteration, with the initial $\psi_\new^{[0]} = \psi \cup \{\psi'\}$ and the output $\psi_\new = \psi_\new^{[k]}$. We claim the following loop invariant: at the end of iteration $i$, for all $h \le i$, $\psi_h \in \psi_\new$ if and only if $\psi_h$ is non-redundant with respect to $\psi_\new^{[i]}$ and $\bigcup \mathcal{C}_\new$.
	
	As inductive hypothesis, suppose the loop invariant holds at the end of iteration $i-1$: for all $h \le i-1$, $\psi_h \in \psi_\new$ if and only if $\psi_h$ is non-redundant w.r.t. $\psi_\new^{[i-1]}$ and $\bigcup \mathcal{C}_\new$. During iteration $i$, $\psi_i$ is removed if and only if the condition in line~\ref{alg:MinimalRefinement-line:RedundancyTest} holds; by Lemma~\ref{lem:BipartiteGraph} this condition being true corresponds to $\psi_i$ being redundant w.r.t. $\psi_\new^{[i-1]}$. Hence, $\psi_i$ is not removed if and only if it is not redundant w.r.t. $\psi_\new^{[i-1]}$.
	
	Moreover, $\psi_\new^{[i]} \subseteq \psi_\new^{[i-1]}$, since at most one removal occurs during an iteration of the loop. Therefore, by Lemma~\ref{lem:MonotonicityNonRedundancy} $\psi_i$ is not removed if and only if it is not redundant w.r.t. $\psi_\new^{[i]}$. This completes the inductive step: we have proved that for all $h \le i$, $\psi_h \in \psi_\new$ if and only if $\psi_h$ is non-redundant with respect to $\psi_\new^{[i]}$.
	
	The property trivially holds for $i = 1$. In this case $\psi_1$ is removed from $\psi_\new^{[0]} = \psi$ if and only if it is redundant w.r.t. $\psi$. And again by lemma~\ref{lem:MonotonicityNonRedundancy} this implies that it is kept in $\psi_\new^{[1]}$ if and only if it is non-redundant w.r.t. $\psi_\new^{[1]}$.
	
	Hence the loop invariant holds for $i = k$ and every $\psi_i \in \psi_\new$ is non-redundant w.r.t. $\psi_\new$ and $\bigcup \mathcal{C}_\new$. By Definition~\ref{def:MinimalAssumptionsSet}, $\psi_\new$ is minimal w.r.t. $\bigcup \mathcal{C}_\new$.
\end{IEEEproof}

\subsection{Time Complexity}
Executing \textsc{MinimalRefinement} introduces an additional overhead to the generation of each refinement compared with the state-of-the-art BFS strategy. In the following we determine the execution time of a single call to this function.

Let $m_\mathcal{C}$ be the number of counterstrategies in $\bigcup \mathcal{C}$ and $m_\psi$ the number of assumptions in $\psi$, and $|Q^{max}_\counterstrategy|$ the maximum number of states in a counterstrategy. Executing lines~\ref{alg:MinimalRefinement-line:StartComputeCPrime}-\ref{alg:MinimalRefinement-line:EndComputeCPrime} requires $m_\mathcal{C}$ LTL model checking operations; each of these involves a counterstrategy and a single GR(1) assumption, which may be either an initial condition, an invariant, or a fairness condition. When using the algorithm in \cite{Clarke1994}, LTL model checking is converted into a problem of Computation Tree Logic (CTL) model checking with fairness conditions: a fairness condition is generated for each $\always$ and $\eventually$ subformula in the formula to check; the full reduction procedure is described in the cited paper. Since all checked formulae contain at most two of these operators, there are at most two fairness conditions in the CTL problem. Solving this problem with a symbolic algorithm requires applying two nested alternating fixpoint operations for each fairness condition (chapter 6 of \cite{Clarke2000}), each taking a number of iterations upper bounded by the number of states in the model to check (see also \cite{Browne1997} regarding complexity of nested fixpoint operations). In summary, solving one of the model checking problems requires $O(|Q^{max}_\counterstrategy|^2)$, and since $m_C$ model checks are performed, the total computation time of lines~\ref{alg:MinimalRefinement-line:StartComputeCPrime}-\ref{alg:MinimalRefinement-line:EndComputeCPrime} is $O(m_C|Q^{max}_\counterstrategy|^2)$.

The bipartite graph in lines~\ref{alg:MinimalRefinement-line:StartMinRefComputation}-\ref{alg:MinimalRefinement-line:EndMinRefComputation} has size $O(m_\psi m_\mathcal{C})$: it requires this amount of computation to be built, and asymptotically as many operations to be minimized. Also notice that in the worst case Algorithm~\ref{alg:MinimalRefinement} produces as many assumptions as counterstrategies in a refinement: so, $m_\psi \le m_\mathcal{C}$, yielding an execution time of $O(m_\mathcal{C}^2)$ for the graph operations.

In summary, executing \textsc{MinimalRefinement} requires an asymptotic time of $T_{MinRef} = O(m_C|Q^{max}_\counterstrategy|^2 + m_\mathcal{C}^2)$. Let us compare this complexity with the time $O(nm_\psi |Q|^2)$ for realizability checks and counterstrategy computation (lines~\ref{alg:ModifiedRefinement-line:RealizabilityTest}-\ref{alg:ModifiedRefinement-line:ComputeCounterstrategy} of Algorithm~\ref{alg:RefinementSearch}). If assumption minimization was not performed, then $m_\psi = m_\mathcal{C}$ for every refinement, and counterstrategy computation would take $O(m_\mathcal{C} |Q|^2)$. By introducing minimization, a part of this contribution quadratic in $|Q|$ is replaced by using \textsc{MinimalRefinement}, quadratic in $|Q^{max}_\counterstrategy|$; typically counterstrategies contain significantly fewer states than entire games (which grow exponentially with the number of variables in the system).

For each explored refinement, the minimization function is called as many times as assumptions returned by \textsc{ApplyBias}; let us denote this number as $|\Psi'|$. Existing generation methods yield a $|\Psi'|$ in the order of tenths \cite{CavezzaA16}. Putting all together, for small $m_\mathcal{C}$, such that the quadratic term in $T_{MinRef}$ is not dominating, we obtain a speedup factor proportional to
$$\frac{m_\mathcal{C} |Q|^2}{m_\psi |Q|^2 + |\Psi| m_\mathcal{C} |Q^{max}_\counterstrategy|^2} \approx \frac{m_\mathcal{C}}{m_\psi}$$
for a single refinement exploration. This can be interpreted this way: minimizing refinements yields the same gain in computation time as one would obtain by just exploring shorter nodes; the overhead induced by the actual minimization operations is negligible.

\end{document}